\documentstyle[epsfig]{aipproc}

\newcommand{\gr}{$\gamma$-ray \,}
\newcommand{\grs}{gamma rays}

\begin{document}

\title{The Diffusive Galactic GeV/TeV Gamma-Ray 
Background:\\ Sources vs. Transport }

\author{Heinrich J. V\"olk$^*$}

\address{$^*$ Max-Planck-Institut f\"ur Kernphysik\\
P.O. Box 103980, 69029 Heidelberg, Germany}


\maketitle

\begin{abstract}
The diffuse Galactic \gr background, as observed with EGRET on CGRO,
exceeds the model predictions significantly above 1 GeV. This
is particularly true for the inner Galaxy. We shall discuss here the
contribution of the Galactic Cosmic Ray (GCR) sources, considered as
unresolved, and in addition the possibility that the transport of the GCRs
out of the Galaxy is not uniform over the Galactic disk. In both cases the
spectrum of the diffuse \grs is harder than the GCR spectrum in the
neighborhood of the Solar system, as observed in situ. The source
contribution is a necessary and, as it turns out, significant part of the
diffuse background, whereas the transport effect is one of several
conceivable additional causes for the hard diffuse \gr spectrum observed.
\end{abstract}

\section*{Introduction} The observations of the diffuse Galactic \gr
emission can be described rather well by a suitable model for the diffuse
interstellar gas, GCR, and photon distributions (e.g. Hunter et al.
1997a). However, above 1~GeV the observed average diffuse \gr intensity,
foremost in the inner Galaxy, $300^{\circ}<l<60^{\circ}$, $|b|\leq
10^{\circ}$, exceeds the model prediction significantly. As far as the
energetic particles are concerned, there are at least two possible
explanations for this discrepancy (e.g. Weekes et al. 1997; Hunter et al.
1997b, and references therein). The high-energy $\gamma$-ray excess may
indicate that the GCR spectrum observed in the local neighborhood is not
representative of the diffuse CR population in the Galactic disk. An
unresolved distribution of CR sources is the other possibility. Since the
\gr emission is the product of the energetic particle intensity on the one
hand, and of the gas density or the photon density, on the other, it is of
course possible that deviations from the above model assumptions for these
latter densities across the Galaxy can also lead
to
changes in the observed energy spectrum of the diffuse \grs. We shall not
discuss such deviations here. We shall rather evaluate the contribution of
the sources, assumed to be the ensemble of Supernova Remnant (SNR) shells,
following a recent calculation by Berezhko \& V\"olk (1999). We shall also 
consider
the transport of the particles from the same sources out of the Galaxy to
naturally increase with decreasing Galactic radius (Breitschwerdt et al.,
1991). We shall leave aside the possibility of new
sources of particles, not known in the neighborhood of the Solar system.

\section*{Gamma rays from the ensemble of SNRs}
Since at best a handful of shell SNRs could be argued to have been
detected up to now in \grs, we shall ignore their discrete contributions
and consider the CR sources to be spatially averaged over the volume
$V_g=2.5\times 10^{66}$~cm$^3$ of the Galactic gas disk, with a radius of
10~kpc and a thickness of 240~pc. The corresponding gas mass is
$M_g=4\times 10^9M_{\odot}$ (Dickey \& Lockman, 1990). The source input
rate in the form of energetic particle energy equals $\nu_{SN} \, \delta
\, E_{SN}$, where we take $\nu_{SN}=1/30~yr$, $E_{SN}=10^{51}$ erg. The
efficiency per SNR is $\delta <1$. The total number of localized SNRs
which still contain their shock accelerated CRs, called here the source
CRs (SCRs), is given by $N_{SN}=\nu_{SN}T_{SN}$, where $T_{SN}$ is their
assumed life time, i.e. the time until which they can confine the
accelerated particles in their interior. Thus $N_{SN}$ is dominated by the
population of old SNRs. We estimate $T_{SN} \simeq 10^5$ yr. After the
time $T_{SN}$ the SCRs rather quickly become part of the ordinary GCRs
that presumably occupy a large Galactic residence volume uniformly.

\subsection*{Acceleration model}
We assume the overall SCR number inside a single SNR to be given by a
power law spectrum $N_{SCR} dE \propto \epsilon^{-\gamma_{SCR}}~dE$ in
energy $\epsilon$ in the relativistic range.

Averaged over the disk volume, the spatial density $n_{SCR}(\epsilon)$ of
SCRs is given $n_{SCR}(\epsilon)=N_{SCR}(\epsilon)N_{SN}/V_g$,
with energy density $e_{SCR}=N_{SN}\delta E_{SN}/V_g$. In terms of
$e_{SCR}$, we have 
\begin{equation} n_{SCR}(\epsilon)=\frac{n_0^{SCR}(\gamma_{SCR}-1)}{mc^2}
 \left( \frac{\epsilon}{mc^2}\right)^{-\gamma_{SCR}}
\end{equation}
and 
\begin{equation}
n_0^{SCR}=\frac{(\gamma_{SCR} -2)e_{SCR}}{(\gamma_{SCR}-1)mc^2},
\end{equation}
for $\gamma_{SCR}>2$. The same expressions hold for the GCRs, given
$e_{GCR}$ and $\gamma_{GCR}$ .

For the SCR we may quite possibly have $\gamma_{SCR}=2$, and then
\begin{equation} n_0^{SCR}=\frac{e_{SCR}}{mc^2\ln(\epsilon_{max}/mc^2)},
\end{equation}
where $\epsilon_{max}\simeq 10^5~mc^2$ is the maximum SCR energy.

The $\pi^0$-decay prodution rate is given by 
\begin{equation}
Q_{\gamma}(\epsilon)=Z_{\gamma}\sigma_{pp} c N_g n(\epsilon),
\end{equation}
(Drury et al. 1994), which leads to the ratio
$R=Q_{\gamma}^{SCR}/Q_{\gamma}^{GCR}$ of the $\gamma$-ray production rates
due to SCRs and GCRs, given by
\begin{eqnarray} R(\epsilon_{\gamma})&=&
\frac{Z_{\gamma}^{SCR} N_{SN}\delta E_{SN}}
{Z_{\gamma}^{GCR}(\gamma_{GCR}-2) \ln
(\epsilon_{max}/mc^2) V_g e_{GCR}}\nonumber \\
             & \times &
\zeta \left(\frac{\epsilon_{\gamma}}{mc^2}\right)^{\gamma_{GCR}-2},
\end{eqnarray}
where $\zeta$ is the ratio $N_g^{SCR}/N_g^{GCR}$, $N_g^{SCR}$ is the mean
source gas density, and $N_g^{GCR}$ denotes
the average gas density in the disk.

With $\delta =0.2$, $e_{GCR} \simeq 2 \times 10^{-12}$~erg/cm$^3$ for the 
relativistic part of the GCRs, and   
$\gamma_{GCR}=2.75$ which results in
$Z_{\gamma}^{SCR}/Z_{\gamma}^{GCR}=10$ (Drury et al. 1994), we obtain 
\begin{equation} R(\epsilon_{\gamma})=0.16
\zeta \left(\frac{T_{SN}}{10^5~\mbox{yr}}\right)
\left(\frac{\epsilon_{\gamma}}{1~\mbox{GeV}}\right)^{0.75},
\end{equation}
for $\gamma_{SCR}=2$.

The total \gr spectrum measured from an arbitrary
Galactic disk volume is then expected to be
\begin{equation} {dN^{\gamma}
\over d\epsilon_{\gamma}}= {dN^{\gamma}_{GCR}\over d\epsilon_{\gamma}}
[1.4+R(\epsilon_{\gamma})],
\end{equation}
where the additional factor 0.4 is introduced to approximately take into 
account the contribution of GCR electron component to the diffuse \gr 
emission at GeV energies, and where $dN^{\gamma}_{GCR}\over
d\epsilon_{\gamma}$ is taken from the paper by (e.g. Hunter et al. 1997b).

\subsection*{"Leaky Box"-type model}
We can derive very similar results from a leaky box-type balance equation
\begin{equation} \frac{n_{GCR}(\epsilon)}{\tau_c}=
\frac{N_{SCR}(\epsilon)}{V_c(\epsilon)}\nu_{SN},
\end{equation}
where $V_c(\epsilon)$ is the energy-dependent residence volume occupied by
GCRs that reach the gas disk during their constant mean residence time
$\tau_c
\simeq 3\times 10^7$~yrs in $V_c(\epsilon)$. In the case of an
extended Galactic Halo, $V_c(\epsilon\gg 1{\rm GeV})\gg V_{g}$ (Ptuskin et
al.
1997). Using eq. (4) we can write 
\begin{equation}
\frac{n_{SCR}}{n_{GCR}}=\frac{V_cT_{SN}}{V_g\tau_c} =
\frac{T_{SN}}{\tau_g}.
\end{equation}
The GCR residence time in the disk
volume 
\begin{equation}
\tau_g=\tau_c V_g/V_c=\frac{x V_g}{v M_g}
\end{equation}
can be derived from the measured grammage 
$x=14\, v/c\, (\epsilon/4.4~\mbox{GeV})^{-0.60}$~g/cm$^2$, for
$\epsilon>4.4$ GeV, and $x=14 v/c$~g/cm$^2$, for $\epsilon<4.4$ GeV
(Engelman et al. 1990).

At relativistic energies $\epsilon > mc^2$, the GCR spectrum and the
overall SCR spectrum $N_{SCR}\propto \epsilon^{-\gamma_{SCR}'}$ are
connected by the relation
\begin{equation} \gamma_{SCR}'=\gamma_{GCR}-0.6=2.15.
\end{equation}
Taking $\gamma_{SCR}=2.15$, which leads to
$Z_{\gamma}^{SCR}/Z_{\gamma}^{GCR}=7.5$ (Drury et al. 1994), we
obtain for $\epsilon_{\gamma}\geq 4.4$~GeV:
\begin{equation}
R(\epsilon_{\gamma})=0.06 \zeta
\left(\frac{T_{SN}}{10^5~\mbox{yr}}\right)
\left(\frac{\epsilon_{\gamma}}{1~\mbox{GeV}}\right)^{0.6}
\end{equation}
(Berezhko \& V\"olk, 1999).

The question is, of course, whether the SN confinement time $T_{SN}$ is
time dependent. Probably this dependence is  
$T_{SN}(\epsilon)=t_0({\epsilon}/{\epsilon_{max}})^{-5}$, where $t_0$ is
the sweep-up time when the SNR enters the Sedov phase and the shock speed
begins
to decrease with time. For average ISM parameters $t_0\sim 10^3$ yr.

\begin{figure}[t] 
\centerline{\epsfig{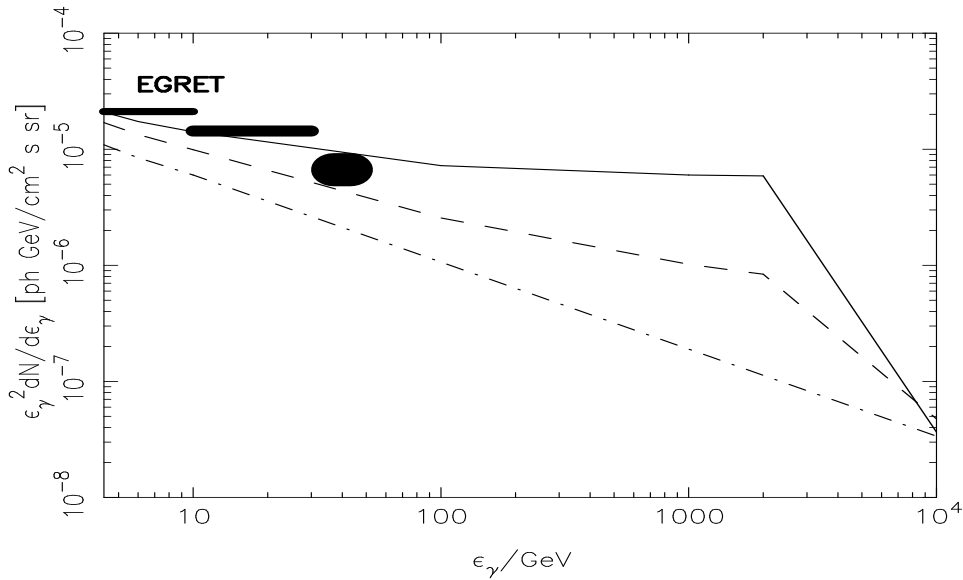}}
\vspace{10pt}
\caption{The differential diffuse \gr energy flux vs \gr energy above 4.4
GeV (cf. Berezhko \& V\"olk, 1999). The heavy symbols are the EGRET
measurements, and the dash-dot line
is the model prediction of Hunter et al. (1997a). The full curve
corresponds to our acceleration model with $\gamma_{SCR}=2$, whereas the  
dashed curve
corresponds to the Leaky Box model. Both theoretical curves incorporate
energy-dependent loss from the acceleration region.}
\label{fig1}
\end{figure}

\section*{Results including the SCRs}
In Fig.~1 we show the measurements by Hunter et al. (1997a) and our
two estimates for the total \gr emission, from GCRs plus SCRs. They
demonstrate
that the SCR contribution for the acceleration model exceeds the leaky box
values for all energies. The reason is that for our empirical model the
acceleration efficiency for the relativistic part of the spectrum is only
$\delta \simeq 0.08$. This is probably due to the fact that the mean
injection efficiency at the SNR shock is lower than the values typically
assumed for a parallel shock by a factor of a few. We take the lower value
for the \gr emission in Fig.1 as the most reliable estimate for the
expected diffuse \gr emission, including the SCRs. Nevertheless the SCR
distribution, which is about 10 percent at GeV energies, becomes dominant
beyond 100 GeV, and exceeds the GCR emission at 1 TeV by almost a factor
of 10. It would be very interesting to detect the diffuse Galactic \gr
emission at 1 TeV in order to test this prediction.

Until now we have only discussed the \gr emission from hadronic SCRs. In
fact, there are many reasons to assume that electrons are equally well
accelerated in SNRs, even if their injection into the shock acceleration
process is much less well understood. The inverse Compton emission by SCR
electrons can be comparable with the hadronic emission, even though it
does not contribute at TeV energies. For a more detailed discussion we
refer to the paper of Berezhko \& V\"olk (1999).

\section*{Transport effects} 
\begin{figure}[t] 
 \centerline{\epsfig{file=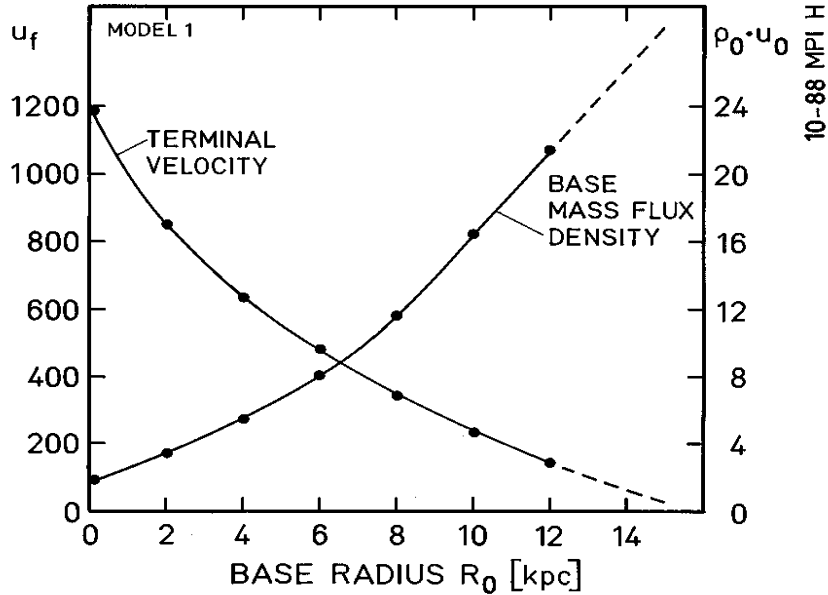,height=4.5in,width=3.5in,angle=90}}  
\vspace{10pt}
\caption{The terminal Galactic Wind velocity, and the base mass flux
density, as functions of radius in the Galactic disk, cf. Breitschwerdt
et al. (1991). All ISM parameters at the base of the wind were considered 
uniform. The radial variation of the Galactic gravitational field alone
is sufficient to produce this radial gradient}
\label{fig2}
\end{figure}

The models used to fit the \gr data from, say, EGRET assume a GCR energy
spectrum that is uniform throughout the Galaxy. This tacitly assumes that
the GCR transport properties leading to the escape from the Galaxy are
everywhere the same. However that needs not be the case, and in fact is
almost certainly not true. The dynamical processes leading to GCR escape
depend on the strength of the regular magnetic field and on its
fluctuation characteristics, as well as on the CR pressure, and the
gravitational field. An example is the formation of Parker bubbles which
remove the enclosed CRs through their boyant rise into the Halo and
ultimately into the Intergalactic Medium. Another example which we wish to
discuss here in some more detail, involves the Galactic Wind which is
partly driven by the GCRs themselves (e.g. Breitschwerdt et al., 1991,
1993; Zirakashvili et al., 1996). In fact, the wind velocity perpendicular
to the disk - in z-direction - is much larger in the central regions of
the Galaxy than at larger radii, through the radial variation of the
Galactic gravitational field alone (see Fig. 2). This implies that for a
given particle energy the boundary seperating the dominantly diffusive
transport perpendicular to the Galactic disk near the disk from the
dominantly convective transport at greater halo heights {\it moves down}
in direction to the Galactic midplane in the inner Galaxy. Since the GCR
diffusion coefficient increases with energy, the position of this boundary
will depend on energy. As shown by Ptuskin et al. (1997), the energy
spectrum of the GCRs is typically $\propto E^{-1.9}$ in the convection
region compared to the standard spectrum $\propto E^{-2.7}$ in the
diffusive confinement region of volume $V_c$ discussed in subsecion 2.1. A
line of site that intersects this boundary will therefore receive \grs~
 from two regions of very different GCR energy spectra, emitting
correspondingly harder spectra than does the diffusive confinement region
alone. Qualitatively this implies a hardening of the truly diffuse \gr
spectrum with Galactic longitude towards the inner Galaxy, for given
latitude. However, the effect will disappear for high enough energies when
the convective zone does no more extend into regions of significant gas
density.

Thus, in contrast to the contribution of the sources, this transport
effect looses importance at high energies.

It remains to work out this effect quantitatively. But its very existence
illustrates the interest we should attach to the measurements of the
diffuse Galactic \gr emission over an as wide as possible range of
energies.

\section*{Conclusions}
The foregoing discussion shows that there are at least two
mechanisms of basic physical interest that contribute to a deviation of
the diffuse Galactic \gr emission spectrum from what would be expected
from CR observations in the Solar vicinity. The contribution from the SCRs
is an inevitable one and is essentially sufficient to explain the data at
least for the inner Galaxy; it should be part of the \gr
emission model to begin with. Clearly this does not rule out effects from
potentially existing new populations of CRs, especially electrons, or the
influence of an increased strength, for instance, of the Interstellar
radiation field. This is particularly true for high Galactic latitudes. 

{\bf Acknowledgements} The work on the role of the CR sources, summarized
in the first part of the paper, has been done jointly with E.G. Berezhhko.
I am also indebted to V.S. Ptuskin for a discussion on the effects of the
Galactic Wind on the diffuse \gr emission.

\end{document}